\documentstyle[epsfig]{elsart}
\journal{Physics Letters B}
\newcommand{\mrm}[1]{\mbox{\rm #1}}

\newcommand{\rfn}[1]{(\ref{#1})}
\newcommand{\bra}[1]{\left\langle #1\right|}
\newcommand{\ket}[1]{\left| #1\right\rangle}

\newcommand{\beq}{\begin{equation}}
\newcommand{\eeq}{\end{equation}}
\newcommand{\bea}{\begin{eqnarray}}
\newcommand{\eea}{\end{eqnarray}}
\newcommand{\bes}{\begin{eqnarray*}}
\newcommand{\ees}{\end{eqnarray*}}

\newcommand{\nn}{\nonumber}
\newcommand{\Eq}[1]{Eq.~(\ref{#1})}
\newcommand{\meg}{\mu\rightarrow e\gamma}
\newcommand{\mec}{$\mu$--$e$}

\newcommand{\ea}{{\it et al.}}

\newcommand{\np}[1]{Nucl. Phys. {\bf #1}}
\newcommand{\pl}[1]{Phys. Lett. {\bf #1}}
\newcommand{\pr}[1]{Phys. Rev. {\bf #1}}
\newcommand{\prl}[1]{Phys. Rev. Lett. {\bf #1}}
\newcommand{\zp}[1]{Z. Phys. {\bf #1}}
\newcommand{\prep}[1]{Phys. Rep. {\bf #1}}

\newcommand{\mpl}[1]{Mod. Phys. Lett. {\bf #1}}

\def\lsim{\mathrel{\vcenter{\hbox{$<$}\nointerlineskip\hbox{$\sim$}}}}

\begin{document}

\begin{frontmatter}

\hfill FTUV/97-45


\hfill HIP-1997-68/TH

\title{New constraints on $R$-parity violation from \\ $\mu$--$e$ 
conversion in nuclei}
\author[a1]{K. Huitu\thanksref{kati}}
\author[a2]{J. Maalampi\thanksref{jukka}}
\author[a3]{M. Raidal\thanksref{martti}} and 
\author[a3]{A. Santamaria\thanksref{arcadi}}
\address[a1]{Helsinki Institute of Physics, P.O. Box 9, 
FIN-00014 University of Helsinki, Finland}
\address[a2]{Theoretical Physics Division, Department of Physics,
P.O. Box 9, FIN-00014 University of Helsinki, Finland}
\address[a3]{Departament de F\'{\i}sica Te\`orica, IFIC, 
CSIC-Universitat de Val\`encia,\\ E-46100 Burjassot, Val\'encia, Spain} 
\thanks[kati]{E-mail: katri.huitu@helsinki.fi}
\thanks[jukka]{E-mail: maalampi@rock.helsinki.fi}
\thanks[martti]{E-mail: raidal@titan.ific.uv.es}
\thanks[arcadi]{E-mail: Arcadi.Santamaria@uv.es}

\begin{abstract}
We derive new constraints on the products of explicitly $R$-parity violating 
couplings $\lambda$ and $\lambda'$ in MSSM from  searches for $\mu$--$e$ 
conversion in nuclei. We concentrate on the loop induced photonic 
coherent conversion mode. For the combinations $|\lambda\lambda|$
which in \mec\ conversion can be probed only  
at loop level our constraints are in many cases more stringent than 
the previous ones due to the enhancement of the process by large 
$\ln(m^2_f/m^2_{\tilde f}).$  For the combinations of $|\lambda'\lambda'|$
the tree-level \mec\ conversion constraints are usually more 
restrictive than the loop ones 
except for two cases which involve the third generation. 
With the expected improvements in the experimental sensitivity, the \mec\
conversion will become the most stringent test for  {\it all} the involved
combinations of couplings. 
\end{abstract}
\end{frontmatter}

\section{Introduction}

The minimal supersymmetric standard model (MSSM) extended by explicit
breaking of $R$-parity, $R=(-1)^{3(B-L)+2S}$ \cite{FF},
 has recently received a lot of attention. 
In MSSM, the conservation of $R$-parity is put in by hand.
Therefore, it is theoretically not well motivated.
On the other hand, models exist in which the violation of
$R$-parity is a necessity, e.g.,  the supersymmetric left-right model
\cite{KM}. Furthermore,
the observed excess of high $Q^2$ events in HERA, if interpreted
in the framework of supersymmetry, seems to indicate
non-zero $R$-parity 
violating  couplings  of squarks \cite{altarelli1}.

Limits on $R$-parity violating  couplings have been derived 
from a large variety of  
observables. Since the 
presence of $R$-parity violation leads automatically to the presence of lepton
and/or baryon number violating processes, searches for them
are especially suitable for constraining these couplings.
In this Letter we  derive new upper limits on the products of the couplings
$\lambda_{ijk}$ and $\lambda'_{ijk}$ from the non-observation of 
\mec\ conversion in nuclei occuring at one-loop level. We compare our
results with the previously obtained constraints including the ones
derived from the tree-level \mec\ conversion.
The latter process
provides particularly restrictive bounds on some products of the couplings
 \cite{kim}, but not for all of them.
The loop induced conversion although 
suppressed by the loop factors is partly sensitive to different couplings than
the tree-level process. Moreover, it was pointed out  in Ref. \cite{RS} that
in a wide class of models with  effective interactions of four charged 
fermions,
thus including the MSSM extended by $R$-parity violation, the 
\mec\ conversion at one-loop is enhanced by large logarithms. 
Therefore, as will be seen,
  the  \mec\ conversion at loop level is more sensitive  
to certain combinations of couplings than the tree-level process.

The motivation for the present work is twofold. Firstly, our study enables us
to put more restrictive bounds on some combinations of the couplings than 
do the previous works. Secondly, the expected improvements in the sensitivity 
of the \mec\ conversion experiments running presently at  PSI  will make 
$\mu$--$e$ conversion the main probe of muon flavour  conservation for
most of the extensions of the standard model allowing for an order of magnitude
more stringent tests of $R$-parity violation already in forthcoming months. 
Indeed, while the present bound on the conversion branching ratio in 
$_{22}^{48}Ti$ is 
$R^{Ti}_{\mu e}\lsim 4.3\cdot 10^{-12}$ \cite{psiti1}, the
SINDRUM II experiment 
taking currently data on gold, $_{79}^{179}Au,$ should reach 
 $R^{Au}_{\mu e}\lsim 5\cdot 10^{-13}$  and next year on titanium
$R^{Ti}_{\mu e}\lsim 3\cdot 10^{-14}$ \cite{psiti2}. 
Furthermore, very recently a proposal for an experiment called MECO 
has been submitted to BNL \cite{meco} which is planned to achieve 
the sensitivity better than $10^{-16}.$ The experimental prospects 
for searches of muon number violation  
have been recently reviewed by A. Czarnecki \cite{czarnecki}.

\section{Previous constraints}

To compare our analyses with the previous ones,
 we first collect and list the updated upper bounds on
all products of  $R$-parity violating couplings which are testable in
\mec\ conversion.
Within the MSSM particle content  the gauge invariance and supersymmetry 
allow for the following $R$-parity violating superpotential \cite{AR}
\beq
W_{R\!\!\!\!/} = \lambda_{ijk} \widehat L_i\widehat L_j\widehat E^c_k +
\lambda '_{ijk} \widehat L_i\widehat Q_j\widehat D^c_k +
\lambda ''_{ijk} \widehat U^c_i\widehat D^c_j\widehat D^c_k
-\mu_i L_i H_2\,,
\label{WRviol}
\eeq
where $\lambda_{ijk} = -\lambda_{jik} $ and 
$\lambda ''_{ijk} = -\lambda ''_{ikj} $.
The $\lambda,\,\lambda '$ and $\mu  $ terms
violate the lepton number, whereas the $\lambda '' $ terms
violate the baryon number by one unit.
The last bilinear term in \Eq{WRviol} gives rise to interesting 
physics which has been studied elsewhere \cite{valle}. 
We will not consider these effects in the 
present paper and
we take as our working model the MSSM extended by the 
trilinear $R$-parity violating terms in \Eq{WRviol}.

Simultaneous presence of $L$ violating ($\lambda_{ijk}, 
\lambda '_{ijk} $) and $B$ violating ($\lambda ''_{ijk} $) couplings
would  lead to too fast proton decay unless
 $|\lambda '\cdot\lambda ''| < 10^{-9}$ \cite{SV}
for squark masses below 1 TeV.
Several examples exist of models, in which there are 
huge differences between the strengths of lepton and baryon number
violating couplings.
In models, which break $R$-parity spontaneously,
only the lepton number is violated.
There are also 
examples of GUT models, in which quarks and leptons are treated
differently \cite{HS,Hempfling}. In the following we assume that only 
lepton number violating couplings are non-vanishing.

\begin{table}
\begin{tabular}{|c|c|c|c|}
\hline \hline
& previous bounds & \multicolumn{2}{c|}{\mec\ at loop level$/\sqrt{B}$} \\
\cline{2-4}
& $m_{\tilde f}=$ & \multicolumn{2}{c|}{$m_{\tilde f}=$} \\
&  100 GeV & $\,$ 100 GeV $\,$ &$\,\;$ 1 TeV $\,\;$  \\ 
\hline \hline
$|\lambda_{121}\, \lambda_{122}|$ & $6.6\cdot 10^{-7}$ \cite{CR} &
$4.2\cdot 10^{-6}$ & $3.2\cdot 10^{-4}$ \\
$|\lambda_{131}\, \lambda_{132}|$ & $ 6.6\cdot 10^{-7}$ \cite{CR} &
$5.3\cdot 10^{-6}$ & $3.9\cdot 10^{-4}$ \\
$|\lambda_{231}\, \lambda_{232}|$ & $ 5.7\cdot 10^{-5}$ \cite{CH} &
 $5.3\cdot 10^{-6}$ & $3.9\cdot 10^{-4}$ \\
$|\lambda_{231}\, \lambda_{131}|$ & $ 6.6\cdot 10^{-7}$ \cite{CR} & 
$8.4\cdot 10^{-6}$ & $6.4\cdot 10^{-4}$ \\
$|\lambda_{232}\, \lambda_{132}|$ & $ 1.1\cdot 10^{-4}$ \cite{CH} &
$8.4\cdot 10^{-6}$ & $6.4\cdot 10^{-4}$ \\
$|\lambda_{233}\, \lambda_{133}|$ & $ 1.1\cdot 10^{-4}$ \cite{CH} & 
$1.7\cdot 10^{-5}$ & $1.0\cdot 10^{-3}$ \\
\hline \hline
\end{tabular}
\caption{
Upper limits on the products  $|\lambda\lambda| $ testable in \mec\
conversion  for two different 
scalar masses $m_{\tilde f}=$100 GeV and  $m_{\tilde f}=$1 TeV.
The previous bounds scale quadratically with the sfermion mass.
The scaling factor $B$ is defined in the text and currently $B=1\,.$}
\end{table}

Severe limitations come also from baryogenesis considerations if
one requires that the GUT scale baryon asymmetry is preserved.
It has been shown in
\cite{DR} that any bound from cosmology can be avoided
by demanding that one of the lepton numbers is 
conserved over cosmological time scales.
Since we wish to study the \mec\ conversion,
we could have conservation of $\tau $ lepton number.
On the other hand, if electroweak baryogenesis is assumed (see,
e.g. \cite{FY}), the restrictions on lepton number are removed.
Therefore we do not impose any extra constraints on lepton
number violating couplings.

\begin{table}
\begin{tabular}{|c|c|c|c|c|}
\hline \hline
 & previous bounds  &
\mec\ at tree-level$/\,\sqrt{B}$ &
\multicolumn{2}{c|}{\mec\ at loop level$/\,\sqrt{B}$} \\
\cline{2-5}
 & $m_{\tilde f}=$ & $m_{\tilde f}=$ &
\multicolumn{2}{c|}{$m_{\tilde f}=$}  \\
& 100 GeV & 100 GeV & $$ 100 GeV $$ &  $\;$ 1 TeV $\;$  \\ 
\hline \hline
$|\lambda '_{211}\, \lambda '_{111}|$ 
 &$5.0\cdot 10^{-6}$ &
$8.0\cdot 10^{-8}$&$1.1\cdot 10^{-5}$ &$8.3\cdot 10^{-4}$  \\
$|\lambda '_{212}\, \lambda '_{112}|$ 
 &$1.4\cdot 10^{-4}$ &
$8.5\cdot 10^{-8}$ &$1.1\cdot 10^{-5}$ &$8.3\cdot 10^{-4}$ \\
$|\lambda '_{213}\, \lambda '_{113}|$ 
 &$1.4\cdot 10^{-4}$ & 
$8.5\cdot 10^{-8}$& $1.1\cdot 10^{-5}$ &$8.3\cdot 10^{-4}$ \\
$|\lambda '_{221}\, \lambda '_{111}|$ 
 &$5.0\cdot 10^{-6}$ &
$4.0\cdot 10^{-7}$&$3.7\cdot 10^{-5}$ &$2.5\cdot 10^{-3}$  \\
$|\lambda '_{222}\, \lambda '_{112}|$ 
 &$4.8\cdot 10^{-5}$ \cite{CR} &
$4.3\cdot 10^{-7}$&$3.7\cdot 10^{-5}$ &$2.5\cdot 10^{-3}$  \\
$|\lambda '_{223}\, \lambda '_{113}|$ 
 &$4.8\cdot 10^{-5}$ \cite{CR} &
$4.3\cdot 10^{-7}$ &$3.7\cdot 10^{-5}$ &$2.5\cdot 10^{-3}$ \\
$|\lambda '_{231}\, \lambda '_{111}|$ 
 &$8.8\cdot 10^{-5}$ & 
$1.0\cdot 10^{-5}$&$1.2\cdot 10^{-3}$ &$8.3\cdot 10^{-2}$  \\
$|\lambda '_{232}\, \lambda '_{112}|$ 
 &$4.8\cdot 10^{-3}$ & 
$1.1\cdot 10^{-5}$& $1.2\cdot 10^{-3}$ &$8.3\cdot 10^{-2}$ \\
$|\lambda '_{233}\, \lambda '_{113}|$ 
 &$4.8\cdot 10^{-3}$ & 
$1.1\cdot 10^{-5}$ &$1.2\cdot 10^{-3}$ &$8.3\cdot 10^{-2}$ \\
$|\lambda '_{211}\, \lambda '_{121}|$ 
 &$4.8\cdot 10^{-5}$ \cite{CR} &
$4.0\cdot 10^{-7}$ &$7.3\cdot 10^{-5}$ &$2.5\cdot 10^{-3}$ \\
$|\lambda '_{212}\, \lambda '_{122}|$ 
 &$4.8\cdot 10^{-5}$ \cite{CR} &
$4.3\cdot 10^{-7}$&$7.3\cdot 10^{-5}$ &$2.5\cdot 10^{-3}$  \\
$|\lambda '_{213}\, \lambda '_{123}|$ 
 &$4.8\cdot 10^{-5}$ \cite{CR} &
$4.3\cdot 10^{-7}$&$7.3\cdot 10^{-5}$ &$2.5\cdot 10^{-3}$  \\
$|\lambda '_{221}\, \lambda '_{121}|$ 
 &$1.4\cdot 10^{-4}$ &
 $8.0\cdot 10^{-8}$&$2.0\cdot 10^{-5}$ &$1.2\cdot 10^{-3}$ \\
$|\lambda '_{222}\, \lambda '_{122}|$ 
&$1.4\cdot 10^{-4}$ &
$2.1\cdot 10^{-6}$& $2.0\cdot 10^{-5}$ &$1.2\cdot 10^{-3}$ \\
$|\lambda '_{223}\, \lambda '_{123}|$ 
 &$1.4\cdot 10^{-4}$ &
$2.1\cdot 10^{-6}$& $2.0\cdot 10^{-5}$ &$1.2\cdot 10^{-3}$ \\
$|\lambda '_{231}\, \lambda '_{121}|$ 
 &$2.6\cdot 10^{-3}$ &
$2.0\cdot 10^{-6}$ &$4.0\cdot 10^{-4}$ &$2.2\cdot 10^{-2}$ \\
$|\lambda '_{232}\, \lambda '_{122}|$ 
 &$4.8\cdot 10^{-3}$ &
$5.3\cdot 10^{-5}$ & $4.0\cdot 10^{-4}$ &$2.2\cdot 10^{-2}$   \\
$|\lambda '_{233}\, \lambda '_{123}|$ 
 &$4.8\cdot 10^{-3}$ &
$5.3\cdot 10^{-5}$ &$4.0\cdot 10^{-4}$ &$2.2\cdot 10^{-2}$   \\
$|\lambda '_{211}\, \lambda '_{131}|$ 
 &$4.2\cdot 10^{-4}$ & 
$1.0\cdot 10^{-5}$ &$1.2\cdot 10^{-3}$ &$8.3\cdot 10^{-2}$  \\
$|\lambda '_{212}\, \lambda '_{132}|$ 
 &$4.0\cdot 10^{-3}$ & 
$1.1\cdot 10^{-5}$&$1.2\cdot 10^{-3}$ &$8.3\cdot 10^{-2}$  \\
$|\lambda '_{213}\, \lambda '_{133}|$ 
 &$1.2\cdot 10^{-5}$ & 
$1.1\cdot 10^{-5}$& $1.2\cdot 10^{-3}$ &$8.3\cdot 10^{-2}$ \\
$|\lambda '_{221}\, \lambda '_{131}|$ 
 &$4.2\cdot 10^{-4}$ &
$2.0\cdot 10^{-6}$ &$4.0\cdot 10^{-4}$ &$2.2\cdot 10^{-2}$ \\
$|\lambda '_{222}\, \lambda '_{132}|$ 
 &$4.0\cdot 10^{-3}$ &
$5.3\cdot 10^{-5}$& $4.0\cdot 10^{-4}$ &$2.2\cdot 10^{-2}$ \\
$|\lambda '_{223}\, \lambda '_{133}|$ 
 &$1.2\cdot 10^{-5}$ &
$5.3\cdot 10^{-5}$&$4.0\cdot 10^{-4}$ &$2.2\cdot 10^{-2}$  \\
$|\lambda '_{231}\, \lambda '_{131}|$ 
 &$2.6\cdot 10^{-3}$ \cite{CH} &
$8.0\cdot 10^{-8}$ &$8.7\cdot 10^{-5}$ &$3.2\cdot 10^{-3}$ \\
$|\lambda '_{232}\, \lambda '_{132}|$ 
 &$2.6\cdot 10^{-3}$ \cite{CH} &
$1.3\cdot 10^{-3}$& $8.7\cdot 10^{-5}$ &$3.2\cdot 10^{-3}$ \\
$|\lambda '_{233}\, \lambda '_{133}|$ 
 &$4.0\cdot 10^{-4}$ & 
$1.3\cdot 10^{-3}$ &$8.7\cdot 10^{-5}$ &$3.2\cdot 10^{-3}$ \\
\hline \hline
\end{tabular}
\caption{
Upper limits on the products  
$|\lambda'\lambda'|$  for two different scalar masses
$m_{\tilde f}=$100 GeV and $m_{\tilde f}=$1 TeV.
The previous bounds and the tree-level bounds scale quadratically with the 
sfermion mass.
The scaling factor $B$ is defined in the text and currently $B=1\,.$}
\end{table}

Upper bounds on the magnitudes of individual couplings  in Eq. (\ref{WRviol})
have been determined 
 from a number of different sources. 
In particular one has considered  charged current universality, 
$e-\mu -\tau$ universality, forward-backward asymmetry,
$\nu_\mu e$ scattering, atomic parity violation \cite{BGH},
neutrinoless double beta decay \cite{HKK}, $\nu $ masses \cite{HS,GRT}, 
$Z$-boson partial width \cite{Bhatta1}, 
$D^+,$ $\tau$-decays \cite{Bhatta2},  $D^0-\overline{D}^0$ mixing,
$K^+,\, t$-quark decays \cite{AG},
$b\bar b$ production at LEP II \cite{EFP}
 and heavy nuclei decay and $n-\overline{n}$ oscillation \cite{BMDH}. 
The bounds from these works have been collected and updated in recent 
reviews \cite{reviews}.

We should also note that very strong constraints on individual 
couplings $\lambda'$ can follow from the bounds on  electric dipole moments 
of fermions \cite{FH}. However, these bounds are qualitatively 
different from the previously mentioned ones, since they depend strongly on the
existence of additional sources of CP-violation in the MSSM and the 
strength of the additional CP-violation. In fact, the upper 
bounds obtained from electric dipole moments apply for 
$|\lambda'|^2 |A|\sin\phi,$ where $|A|$ is the so-called $A$-term and
$\phi$ is its phase, rather than for $\lambda'.$ If one makes the
assumption that $A$ is real and the only source of CP-violation 
in the MSSM is the CKM matrix then these limits are identically washed away. 
Since  these bounds are very model dependent and easily avoidable 
 we shall not use them in our work.

More recently there has been a lot of interest in deriving bounds on the 
products of two such couplings which are stronger than the products of 
known bounds on the individual couplings.
In addition to giving rise to unobserved rare processes these products
turn out to be interesting also from the model building point of view. 
For example, bounds on these products  
constrain considerably models which attribute the origin of $R$-parity
violation to approximate 
$U(1)$ \cite{yuval} or $U(2)$ \cite{bar3} flavour symmetry.
 Products of the type 
$\lambda\lambda$ have been constrained from searches for 
muonium-antimuonium conversion \cite{kim}, $\meg$ \cite{CH} and
heavier lepton decays to three lighter ones \cite{CR}. Combinations of 
 $\lambda\lambda'$ have been bounded from the tree-level \mec\ conversion,
$\tau$- and $\pi$-decays \cite{kim}, from $K\rightarrow l_1\bar l_2$ 
\cite{CR} and from $B\rightarrow l_1\bar l_2$ 
\cite{JKL1}. Upper limits on  $\lambda'\lambda'$-s are found from 
neutrinoless double beta decay \cite{BM},  $\meg$ \cite{CH}\footnote{
The bounds on $\lambda'\lambda'$-s in Ref. \cite{CH} should all be
divided by 3 due to the omission of the colour factor in the calculation.},
 tree-level \mec\ conversion \cite{kim}, $\Delta m_K,$  $\Delta m_B$
\cite{CR}, exotic top decays \cite{YYZ} and non-leptonic \cite{CRS}
and semi-leptonic \cite{JKL2} decays of heavy quark mesons.

We have presented the
combinations of  
$\lambda\lambda$ and $\lambda'\lambda'$  contributing to the \mec\ conversion 
in Table 1 and Table 2, respectively.
The upper limits on these combinations of couplings from the previous works  
are presented in the tables in the first columns. 
For  $\lambda'\lambda'$-s the tree-level \mec\ conversion bounds
are presented separately in the second column of Table 2.
All the couplings are assumed to be real.
If the previously obtained bound on a particular product of couplings 
is stronger than the product of the bounds on the individual couplings 
then we have also presented in the tables the references from which the bound
has been  taken. In the opposite case the references are not given and
can be found for example in Ref. \cite{reviews}. The tree-level \mec\
bounds are all taken  from Ref. \cite{kim} assuming that there are no
cancellations between different contributing terms.   
The bounds are given for two values, 100 GeV and 1 TeV, 
of the relevant supersymmetric scalar mass. If the listed bounds depend
on the mass quadratically, which is the case for the bounds derived
from tree-level processes,
we have presented them for 100 GeV only. Extrapolation to 
1 TeV can be done trivially by scaling the bound by two orders of magnitude.
The quadratic dependence on the mass is also approximately true for the
$\lambda\lambda $-type couplings in the loop-induced $\mu\rightarrow e\gamma$,
but not for the $\lambda '\lambda '$-type couplings involving the third
family.
In this case the upper bound for $m_{\tilde f}= 1$ TeV becomes
$|\lambda '_{23i}\lambda '_{13i}|<1.9\cdot 10^{-2},\; i=1,2$ \cite{CH}.

The most stringent constraint on the products
$|\lambda_{122}\lambda'_{211}|,$  $|\lambda_{132}\lambda'_{311}|,$
$|\lambda_{121}\lambda'_{111}|$ and $|\lambda_{231}\lambda'_{311}|$
follows from the tree-level \mec\ conversion and is 
$4.0\cdot 10^{-8}$ for $m_{\tilde f}=100$ GeV.

\section{Coherent $\mu$--$e$ conversion}

Due to the extraordinary sensitivity of ongoing experiments searching for
\mec\  conversion  in nuclei, it can be expected that in the nearest 
future this process will become the main test of muon flavour conservation
\cite{RS}. To take full advantage of the \mec\ experiments 
in studying possible new physics we present a general formalism
for the coherent \mec\ conversion. In practice only the appropriate 
expressions for the coherent transition are needed, since the coherent 
process is enhanced by the large coherent nuclear charge and 
it dominates over the 
incoherent transition. Indeed, for the nuclei of interest the coherent rate
of the conversion constitutes about 91-94\% of the total rate \cite{chiang}.

The theory of \mec\ conversion in nuclei was first studied by Weinberg 
and Feinberg \cite{WF} and the early developments in this field have been
summarised in  \cite{ver3}. Since then 
various nuclear models and approximations are used in the literature to 
calculate the coherent \mec\ conversion 
 nuclear form factors. It is important to notice that 
the results from the shell model \cite{ver1}, local density approximation
\cite{chiang} as well as quasi-particle RPA approximation \cite{ver2}
do not differ significantly from each other for both 
$_{22}^{48}Ti$ and $_{82}^{208}Pb$
nuclei showing consistency in understanding of nuclear physics involved
\cite{ver2}. We follow the notation of Ref.~\cite{chiang} taking into
account the corrections to the approximation from the exact calculations 
performed in the same work.
The corrections are negligible for $_{22}^{48}Ti$ as the local density 
approximation works better for light nuclei but they are sizable,
of the order of 40\%,  for $_{82}^{208}Pb$ and $_{79}^{179}Au.$

The relevant \mec\ conversion effective Lagrangian can be expressed as
\beq
{\cal L}_{eff} \,=\,
\frac{4\pi\alpha}{q^2} \, j^{\lambda}_{(ph)} J^{(ph)}_{\lambda}
\, + \, \frac{G}{\sqrt{2}}  \sum_{i,m}
\left( j^{\lambda}_{i(non)} J^{i(non)}_{\lambda}\,+\,j_{m(non)} J^{m(non)}
\right)\,,
\label{me}
\eeq
where the first term   describes  the photonic conversion 
and the second set of terms describe  different contributions to the 
 non-photonic conversion mechanisms.
Here $q^2$ denotes the photon momentum transfer, $G$ the effective coupling
constant appropriate for the interactions considered,
$j^{\lambda}$ and $j$ represent 
the leptonic and  $J^{\lambda}$ and
 $J$ hadronic vector and scalar currents, respectively. 
The photonic mechanism is enhanced by small  $q^2$ but can occur only at
loop level and is thus suppressed by loop factors. 
The non-photonic mode is of interest, if the conversion can occur at 
tree-level like in models with non-diagonal $Z'$ \cite{zprim} and
Higgs  \cite{ng1} couplings, 
models with leptoquarks \cite{davidson}, or  if the
loop contributions are enhanced by some other mechanism, e.g., models
with non-decoupling of massive neutrinos \cite{nondec}. 
In models with broken $R$-parity, one may expect that both modes are
important since the photonic mode is enhanced by large logarithms.

Most generally the leptonic current for the photonic mechanism can be
\linebreak parametrised as 
\bea
j^{\lambda}_{(ph)} &=& {\bar e} \left[ \; 
\left (f_{E0} + \gamma_5 f_{M0}\right) \gamma_{\nu}  \left (
g^{\lambda \nu} - \frac {q^{\lambda} q^{\nu}}{q^2}\right )
\right. 
+\left. 
(f_{M1} + \gamma_5 f_{E1})\; i\; \sigma^{\lambda \nu} \frac{q_{\nu}}{m_{\mu}} 
\right] \mu\,,
\label{j1}
\eea
where the form factors $f_{E0}$, $f_{E1}$, $f_{M0}$ and 
$f_{M1}$ can be computed from the underlying theory.
Since the interaction of photon with fermions 
is well known, the hadronic current 
can  in this case be  written down immediately in terms of the nucleon 
spinor ${\bar \Psi}=({\bar p}, {\bar n})$ as \cite{ver3}
\beq
J^{(ph)}_{\lambda}={\bar\Psi}\gamma_{\lambda}\frac{1}{2}(1+\tau_3) \Psi\,,
\eeq
where $\tau_3$ is the isospin Pauli matrix.

We do not consider non-photonic conversion mode in this Letter.
However, since  in the context of the \mec\ conversion 
only the vector currents and the corresponding 
nuclear matrix elements
have been  discussed in the literature (see, e.g., Ref. \cite{zprim})
while most of the models contain also new scalar particles,  we feel
that the following comment is in order.
To calculate  the tree-level \mec\ conversion rate, one has 
to proceed from the quark level to the nucleon level by computing 
 the quark current matrix elements 
$\bra{N}{\bar q}\gamma_\lambda q\ket{N}= G_V^{(q,N)}{\bar N}\gamma_\lambda N$ 
and $\bra{N}{\bar q} q\ket{N}= G_S^{(q,N)}\, {\bar N} N $
for the nucleon $N=$p, n (the coherent
transition rate depends only on the quark vector or scalar couplings).
 Due to the conservation of the vector current
and its coherent character, with the vector charge equal to the quark number,
and assuming strong isospin symmetry, one can determine 
in the limit $q^2\approx 0$ 
$ G_V^{(u,p)}=G_V^{(d,n)}\equiv G_V^{(u)}=2$ and 
$G_V^{(d,p)}=G_V^{(u,n)}\equiv G_V^{(d)}=1.$
However, in the case of the scalar
operator ${\bar q} q $  one has to rely on nucleon models.
Here we point out two extreme situations. In the limit of the fully
non-relativistic quark model, for which ${\bar q}q$ can be basically 
regarded as a number operator,  it is known that 
$G_S\approx G_V.$  This  also determines the upper
bounds for the scalar form factors. The largest 
differences between the form factors are expected to 
occur in the case of fully relativistic models  
 like the MIT bag model. In the latter case one has estimated
the scalar form factors for massless quarks to be \cite{vento} 
$G_S^{(u)}=4/3$ and $ G_S^{(d)}=2/3.$ 
The actual values of the scalar form factors thus 
lay in between these two limiting cases. Therefore the constraints 
derived from the tree-level \mec\ conversion in Ref. \cite{kim}
can be somewhat relaxed due to this uncertainty  but not more than 
a factor of 2/3 or so.

Due to the large mass of muon it is
appropriate to take the customary non-relativistic limit
for the motion of the muon in the muonic atom. 
In this limit the "large" component of the muon wave function factorizes out 
when calculating the coherent conversion rate. 
Therefore, for the photonic mechanism 
the coherent \mec\ conversion  branching ratio 
$R^{ph}_{\mu e}$ can be expressed as 
\beq
R^{ph}_{\mu e}=C\,\frac{8\pi\alpha^2}{q^4}\,p_e E_e \,
\frac{|M(p_e)|^2}{\Gamma_{capt}}\,Z^2\,\xi_0^2\, ,
\eeq
where $E_e$ is the electron energy, $E_e\approx p_e\approx m_\mu$
for this process, $\Gamma_{capt}$ is the total
muon capture rate, $|M(p_e)|^2$ is the nuclear matrix element squared
and 
\beq
\xi_0^2=|f_{E0}+f_{M1}|^2+|f_{E1}+f_{M0}|^2
\label{ksi}
\eeq
shows the process dependence on the photonic form factors.
The expression for  $|M(p_e)|^2$ in the local density approximation,
\beq
|M(q)|^2=\frac{\alpha^3\,m_\mu^3}{\pi}\frac{Z^4_{eff}}{Z}\,
|\overline{F_p}(q)|^2
\, ,
\eeq
the correction factors $C$ to the approximation (compared with the 
exact calculation) as well as  all numerical values of the above
defined quantities for  $^{48}_{22}Ti$ and $^{208}_{82}Pb$  are presented in
Ref. \cite{chiang,chiang2}. The result reads
\beq
R^{ph}_{\mu e}=C\,\frac{8\alpha^5\,m_\mu^5\,Z^4_{eff}\,|\overline{F_p}(p_e)|^2
Z}{\Gamma_{capt}}
\cdot \frac{\xi_0^2}{q^4}\, ,
\label{mecrate}
\eeq
where $C^{Ti}=1.0,$ $C^{Pb}=1.4,$ $Z^{Ti}_{eff}=17.61,$ $Z^{Pb}_{eff}=33.81,$
$\Gamma_{capt}^{Ti}=2.59\cdot 10^6$ s$^{-1},$
$\Gamma_{capt}^{Pb}=1.3\cdot 10^7$ s$^{-1}$ and the proton nuclear 
form factors are $\overline{F_p}^{Ti}(q)=0.55$ and
$\overline{F_p}^{Pb}(q)=0.25.$
Currently SINDRUM II experiment is running on gold, $_{79}^{179}Au,$
 but gold is not explicitly treated in Ref. \cite{chiang}.
However, since $Z^{Au}=79$ and $Z^{Pb}=82$ are so close to each other
then, within errors, all the needed quantities for   $_{79}^{179}Au$
and $^{208}_{82}Pb$ are approximately equal\footnote{We thank 
H.C. Chiang and E. Oset for clarifying us this point.}. 
This result is strongly supported by theoretical calculations and 
experimental measurements of the total muon capture rate of $Pb$ and $Au$
\cite{chiang2}. In the numerical values we use the same experimental and 
theoretical input for both    $_{79}^{179}Au$ and $^{208}_{82}Pb.$

\section{Bounds from the photonic $\mu$--$e$ conversion}

The \mec\ conversion in supersymmetric theories which conserve $R$-parity 
has been studied before in several works \cite{vergados}. They conclude that 
other lepton flavour violating processes like $\meg$ are about an order of 
magnitude more sensitive to the new phenomena than \mec\ conversion.
However, in the case of broken $R$-parity one may  expect the opposite result
due to the  logarithmic enhancement of \mec\
conversion, which is about one  order of magnitude at the amplitude level.  

Addition of Eq. (\ref{WRviol}) to the MSSM superpotential leads to more 
interactions in the model, but the MSSM particle content remains.
The relevant part of the Lagrangian
is found by  standard techniques \cite{HK}
\bea
L_{L\!\!\! /\, ,\lambda, \lambda '}&=&
\lambda_{ijk}\left[\tilde\nu_{iL}\bar e_{kR} e_{jL}
+\tilde e_{jL}\bar e_{kL} \nu_{iL}
+\tilde e_{kR}^*\overline{\nu_{iL}^c} e_{jL}  
-(i\leftrightarrow j)\right]  \nonumber \\
&+&\lambda '_{ijk}\left[ \left(
\tilde\nu_{iL}\bar d_{kR} d_{jL} + \tilde d_{jL}\bar d_{kR} \nu_{iL}
+ \tilde d_{kR}^* \overline{\nu_{iL}^c} d_{jL}
\right)\right. \nn \\
&& \left. - K_{jp}^\dagger \left( 
\tilde e_i \bar d_{kR} u_{pL} +
\tilde u_{pL}\bar d_{kR} e_{iL}
+\tilde d_{kR}^*\overline{e_{iL}^c} u_{pL}\right)\right] +h.c.,
\label{lagr}
\eea
where we have taken into account the flavour-mixing effects in the 
up-quark sector in terms of the CKM matrix elements $K_{jp}.$
The effects of possible non-alignment of fermion and sfermion mass
matrices are severely constrained \cite{CEKLP} and therefore neglected.

\begin{figure}[t]
\leavevmode
\begin{center}
\mbox{\epsfxsize=10.truecm\epsfysize=10.truecm\epsffile{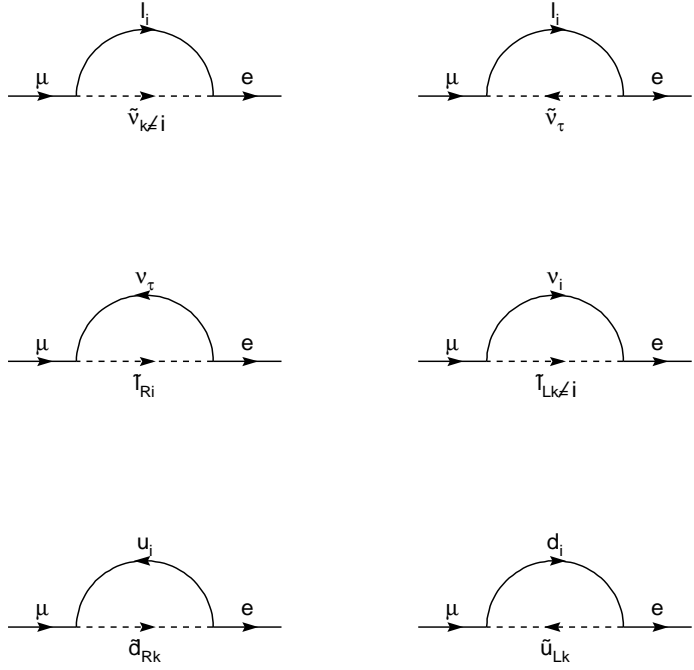}}
\end{center}
\caption{\label{graphs} Contributing loop diagrams for \mec\ conversion.
Unless otherwise indicated $i,k=1\ldots 3$.}
\end{figure}

The Feynman diagrams contributing to the \mec\ conversion at one-loop
are shown in Fig. \ref{graphs}.
The photon line is not shown, but it should be attached in all
possible ways to the graphs.
Following the general effective Lagrangian analyses of Ref. \cite{RS}
we can say immediately that the first two  diagrams with leptons 
in the loop give a logarithmically enhanced contribution to the \mec\
conversion while the second two involving neutrino as an intermediate
state are not enhanced. This is because in the latter set of diagrams 
one cannot attach the photon to the fermion line.
For the typical fermion and sfermion masses we have  
$|\ln (m_f^2/m_{\tilde f}^2)|\sim{\cal O}(10)$ and, therefore, the contribution
of the second set of diagrams can be neglected without affecting our numerical
results. For the rest of the diagrams  we compute
the photonic form factors defined in \Eq{j1}. 
Starting from the Lagrangian \rfn{lagr} and 
using dimensional regularization we obtain for 
the form factors  proportional to $\gamma_\lambda$
\bea
f^{f=\ell}_{E0}=\pm f^{f=\ell}_{M0}&=&
-\;\frac{2 \,(\lambda\lambda)}{3\, (4\pi)^2}\;\frac{-q^2}{m^2_{\tilde f}}
 \left( \ln\frac{-q^2}{m^2_{\tilde f}} + F_f(r)\right)\, , 
\label{fl1} \\
f^{f=u}_{E0}=- f^{f=u}_{M0}&=&
-\;\frac{(\lambda'\lambda')N_c}{9\, (4\pi)^2}\;\frac{-q^2}{m^2_{\tilde f}}
\left(\ln\frac{-q^2}{m^2_{\tilde f}} + F_f(r)-\frac{1}{12}
\right)\, , 
\label{fu1} \\
f^{f=d}_{E0}=- f^{f=d}_{M0}&=&
-\;\frac{(\lambda'\lambda')N_c}{18\, (4\pi)^2}\;\frac{-q^2}{m^2_{\tilde f}}
\left(\ln\frac{-q^2}{m^2_{\tilde f}} +  F_f(r)-\frac{1}{3}
\right)\, ,
\label{fd1} 
\eea
and for the form factors  proportional to $\sigma_{\mu\nu}$
\bea
f^{f=\ell}_{M1}=\mp f^{f=\ell}_{E1}&=&
-\;\frac{(\lambda\lambda)}{3\, (4\pi)^2}\; \frac{m_\mu^2}{m^2_{\tilde f}}\, , 
\label{fl2} \\
f^{f=u}_{M1}=f^{f=u}_{E1}&=&
-\;\frac{(\lambda'\lambda')N_c}{24\, (4\pi)^2}\; 
\frac{m_\mu^2}{m^2_{\tilde f}}\, , 
\label{fu2} \\
f^{f=d}_{M1}= f^{f=d}_{E1}&=& 0\,. 
\label{fd2} 
\eea
Here the superscripts $f=\ell,$ $f=u$ and $f=d$ denote contributions
from the diagrams with leptons, $u$- and $d$-quarks in the loop, 
respectively, $r=m_{f}^2/(-q^2),$  
where  $f$ stands for the virtual
fermion in the loop, $N_c=3$ is the number of colours and
\bea
F_f(r)=-\frac{1}{3}+4 r+\ln r 
 +\left(1-2r \right)
\sqrt{1+4 r} \,\ln\left( 
\frac{\sqrt{1+4r}+1}{\sqrt{1+4r}-1}\right)\, .
\label{f1}
\eea
 There are three important limiting cases. 
If $r\to 0$  then $F_f=-1/3,$ 
if $r\approx 1$  then $F_f\approx 1.52$ 
and if $r\gg 1$  then $F_f=\ln r+4/3.$

As expected the form factors  \rfn{fl1}, \rfn{fu1},  \rfn{fd1} contain 
large logarithms  while the form factors 
 \rfn{fl2}, \rfn{fu2},  \rfn{fd2} do not.
The dependence of the form factors on particular 
combination of $R$-parity violating
couplings as well as on the $\pm$ signs in the form factors coming
from the leptonic loops should be fixed when calculating the constraints 
on the products of couplings.  
Deriving these expressions we have taken into account that $q^2<0$
and made a simplifying  approximation $x_f=m_f^2/m_{\tilde f}^2\ll 1$ 
which allows us to 
keep only the leading terms in  $x_f.$ This approximation is
not adequate only in one case, i.e., if the fermion in the loop is top
quark and $m_{\tilde f}=100$ GeV. However, in this case the \mec\ conversion
is completely dominated by the diagram with the bottom quark 
in the loop and the top 
contribution can  be neglected without making significant numerical 
error.  Note also that in the first approximation \Eq{fd2} vanishes.
For the expressions of $f_{M1,E1}$ in higher orders of $x_f$ we
refer the reader to Ref. \cite{CH}.

Now we are ready to constrain $R$-parity violating couplings.
Substituting the numerical values of the parameters to \Eq{mecrate} 
we find  
\beq
R^{ph}_{\mu e} = 8.6\;(12.3)
\cdot 10^{-4}\;\mrm{TeV}^{4}\;\frac{\xi^2_0}{q^4}
\,,
\label{tirate} 
\eeq
where the first number in the expression corresponds to 
the conversion in $Ti$ and
the number in the brackets to the conversion in $Pb$ and $Au$
and $\xi^2_0$ is defined in \Eq{ksi}.
Note that the conversion is somewhat enhanced in $Pb$ and
$Au$ (in fact, maximised \cite{chiang})  if compared with $Ti.$
Substituting  the current best bound on the conversion branching ratio  
$R^{Ti}_{\mu e}\lsim 4.3\cdot 10^{-12}$ \cite{psiti1} and the 
form factors above to \Eq{tirate} we find the constraints on 
the products of $R$-parity violating couplings from \mec\ conversion
at one-loop level. 
The bounds on the contributing combinations of $\lambda\lambda$
are listed in Table 1 and on the combinations of $\lambda'\lambda'$
in Table 2 for two values of sfermion masses $m_{\tilde f}=100$ GeV and
1 TeV.

In future new experimental data on \mec\ conversion will be available.
To take into account the future improvements in the experimental 
sensitivity we define a scaling factor $B$ as
\beq
B=\frac{R^{Ti}_{\mu e}(present\, bound)}{R^{Nu}_{\mu e}(future\, bound)}
\times \left\{  
\begin{array}{cl}
1.0\,, & \;\;\;\;\mrm{if} \, Nu=Ti\,, \\
2.8\,, & \;\;\;\; \mrm{if}\, Nu=Au\,.
\end{array}
\right.
\eeq
To get new bounds  one just has to divide the bounds in 
Table 1 and Table 2 by the corresponding $\sqrt{B}.$

\section{Discussion and conclusions}

The photonic \mec\ conversion which occurs at one-loop level probes the
coupling products  of the type $\lambda\lambda$ and $\lambda'\lambda'.$
For the combinations of $\lambda\lambda$
comparison with the previously obtained bounds in Table 1 shows that in
three cases out of six our new bounds  are more stringent.
This is because the conversion is enhanced by large logarithms
$\ln (m_f^2/m_{\tilde f}^2).$
 If SINDRUM II experiment will reach the expected sensitivity 
and show negative results one  has to scale
the bounds down  by factors of $\sqrt{B^{Au}}=5$ and  $\sqrt{B^{Ti}}=12$
which correspond to the expected limits 
  $R^{Au}_{\mu e}\lsim 5\cdot 10^{-13}$ and 
 $R^{Ti}_{\mu e}\lsim 3\cdot 10^{-14},$ respectively.
In this case the \mec\ conversion bounds in Table 1
will be more stringent than the previous bounds for {\it all} 
testable combinations of $\lambda\lambda.$
This is an important conclusion since no that significant improvements
of the other experiments is expected.
Note that in \mec\ conversion $\lambda\lambda$-s 
can be probed only at loop level.

The  $\lambda'\lambda'$-type couplings give also rise to the
tree-level \mec\ conversion. The bounds derived from 
this process are very strong 
for  most products of the couplings but not for all of them.
The reason is that for some combinations of the couplings, especially
if the third family squarks are involved, their contribution to the
\mec\ conversion can be strongly suppressed by small off-diagonal
CKM matrix elements.
In the worst case the suppression factor can be  as large as $\lambda_W^6,$
where $\lambda_W\sim 0.2$ is the Wolfenstein parameter. 
As a result the tree-level \mec\ conversion bounds are usually 
more stringent than 
the bounds from the other processes except for three cases.
The bound on $|\lambda'_{223}\lambda'_{133}|$ obtained by multiplying 
the bounds on the individual couplings is stronger than the \mec\ conversion
bound (note also that the previous bound on
$|\lambda'_{213}\lambda'_{133}|$ is only marginally weaker than the 
tree-level \mec\ result).
Importantly, as the result of our calculation, 
the bounds on $|\lambda'_{232}\lambda'_{132}|$ and
$|\lambda'_{233}\lambda'_{133}|$ derived from the {\it loop} induced
\mec\ conversion are  more stringent than the ones from the 
tree-level \mec\ conversion.
With the new data  all the \mec\ conversion bounds on $|\lambda'\lambda'|$-s
will be  more stringent than the others. Since the $B$ factors for 
the loop induced and tree-level conversions are the same  
then the last two entries in Table 2
remain  most stringently constrained by the loop results.

In conclusion, searches for the \mec\ conversion in nuclei,
taking into account both the photonic conversion mode studied in 
this Letter and the non-photonic mode,
constrain most of 
the products of $R$-parity violating couplings testable in this process
more stringently than do the  other processes.  In the nearest future
this conclusion is expected to apply for all the products.

\begin{ack}
M.R. thanks A. van der Schaaf for information about ongoing experiments at PSI
and J. Bernab\'eu, H.C. Chiang, E. Oset and V. Vento 
 for discussions on nuclear physics involved
in $\mu$--$e$ conversion.
This work has been supported in
part by CICYT (Spain) under the grant AEN-96-1718.
\end{ack}

\end{document}